\documentclass{new_tlp}

\usepackage{amsfonts,amsmath}

\submitted{20 January 2022}
\revised{14 March 2022}
\accepted{30 April  2022}

\usepackage{graphicx}
\usepackage{mathptmx}
\usepackage{tikz-qtree}
\usepackage{filecontents}
\usepackage{csvsimple}
\usepackage{amsfonts,amsmath}
\usepackage{subfig} 
\usepackage{url}
\usepackage{verbatim}
\usepackage{color}
\usepackage{amsmath} 
\usepackage{proof}

\newcommand{\imp}{\rightarrow}

\newcommand{\en}{\wedge} 
\newcommand{\of}{\vee} 

\newcommand{\seq}{\Rightarrow}

\newcommand{\De}{\Delta}
\newcommand{\Ga}{\Gamma}
\renewcommand{\phi}{\varphi}

\input prolog.tex

\definecolor{lgray}{gray}{0.95}
\definecolor{lblue}{rgb}{0.90,0.90,1.00}
\definecolor{lyellow}{rgb}{1.00,1.00,0.70}

\usepackage{listings}
\newtheorem{ex}{Example}

\lstnewenvironment{codex}
    {\lstset{}%
      \csname lst@SetFirstLabel\endcsname}
    {\csname lst@SaveFirstLabel\endcsname}
\newcommand{\BI}[0]{\begin{itemize}}
\newcommand{\EI}[0]{\end{itemize}}
\newcommand{\I}[0]{\item}
\newcommand{\BE}[0]{\begin{enumerate}}
\newcommand{\EE}[0]{\end{enumerate}}

\newcommand{\BX}[0]{\begin{ex}}
\newcommand{\EX}[0]{\end{ex}}
\newcommand{\BV}[0]{\begin{verbatim}}
\newcommand{\EV}[0]{\end{verbatim}}

\newcommand{\BF}[0]{\begin{filecontents*}{data.csv}}

\newcommand{\BQ}[0]{\color{blue}\begin{quote}}
\newcommand{\EQ}[0]{\end{quote}\color{black}}

\def \bscale1 {0.25}
\def \bscale {0.25}

\begin{document}

\title{Abductive Reasoning in Intuitionistic Propositional Logic via Theorem Synthesis}
\author{Paul Tarau}
\author[Paul Tarau]
         {
          Paul Tarau\\
          University of North Texas\\
          {\em paul.tarau@unt.edu}\\         
         }

\maketitle

\begin{abstract}
With help of a compact Prolog-based theorem prover for Intuitionistic Propositional Logic, we  synthesize minimal assumptions under which a given formula formula becomes a theorem. 

After applying our synthesis algorithm to cover basic abductive reasoning mechanisms,  we synthesize conjunctions of literals that mimic  rows of truth tables in  classical or intermediate logics and we abduce  conditional hypotheses that turn the theorems of classical or intermediate logics into theorems in intuitionistic logic.
One step further, we generalize our abductive reasoning mechanism to synthesize more expressive sequent premises using a minimal  set of canonical formulas, to which arbitrary formulas in the calculus can be reduced while preserving their provability.

Organized as a self-contained literate Prolog program, the paper supports interactive exploration of its content and ensures full replicability of our results.

\end{abstract}

\section{Introduction}

Given a formula $F$ in Classical Propositional Logic ({\bf CL}), each row in a formula's truth table  describes a conjunction of literals $C$. Reading the truth table as a disjunctive normal form, it immediately follows that $C \rightarrow F$ is a tautology.   
As an example, let us consider the {\bf CL} formula \verb|F = (A v B) & (B v C) & (C v A)|.
Then its truth table (with 1 for True and 0 for False) is the one on the left. Let us select any row, say [1,0,1] and interpret it as \verb|G = A & ~B & C|. Then the truth table of the resulting tautology \verb~G -> F~ is shown on the right.
\begin{codex}
 A B C :  F                                                 A B C:  G->F
[0,0,0]-->0                                                [0,0,0]-->1
[0,0,1]-->0                                                [0,0,1]-->1
[0,1,0]-->0                                                [0,1,0]-->1
[0,1,1]-->1                                                [0,1,1]-->1
[1,0,0]-->0                                                [1,0,0]-->1
[1,0,1]-->1  <= selected row                               [1,0,1]-->1
[1,1,0]-->1                                                [1,1,0]-->1
[1,1,1]-->1                                                [1,1,1]-->1
\end{codex}

Thus, it is easy to extract from the truth-table of a formula $F$ in {\bf CL} assumptions that would make it a theorem in {\bf CL}.

This {\em model-theoretic} approach extends also to intermediate logics\footnote{logics weaker than classical but stronger than intuitionistic} and in particular, it applies to the 5-valued truth-tables of the  equilibrium logic \cite{table5valued}, underlying Answer Set Programming ({\bf ASP}).

With {\em no finite truth-tables}, no inter-definability of logical connectives, no rule of excluded middle and only a concept of {\em tautology} and {\em contradiction} defined for Intuitionistic Propositional Logic ({\bf IL}), we need to be a bit more creative when trying to find  {\em salient} assumptions that would make the formula a theorem in {\bf IL}.  Such salient assumptions include conjunctions of literals, mimicking the truth tables of {\bf CL} but it also makes sense to extend them to more expressive subsets of formulas.
First, given a formula in {\bf IL}, we will need a search process for finding  assumptions that would make it a theorem.  Next, we would like our assumptions to be minimal with respect to the partial order relation governing the logic (or its equivalent Heyting algebra) , {\em intuitionistic implication}.

This brings us to {\em Abductive Logic Programming} \cite{eshghi89,abd_lp}, where facts designated as {\em abducibles} are filtered with integrity constraints to  provide relevant assumptions needed for the success of a goal $G$ w.r.t. a given program $P$. In the context of {\bf IL}, our abductive reasoning will rely on finding {\em minimal assumptions under which a formula becomes a theorem}. 

And finally, in the absence of a convenient automated semantic method like truth tables or SAT solvers in {\bf CL}, we will need a theorem prover, ideally derived directly from the rules of a {\em terminating} sequent calculus, that interoperates smoothly with the search process synthesizing our assumptions.

These requirements make Prolog a natural meta-language for an actionable description of these concepts. We will materialize our approach as a literate Prolog program, from which, as a convenience to the reader, we will extract our code and make it available online as a Prolog file\footnote{at \url{https://github.com/ptarau/TypesAndProofs/blob/master/isynt.pro}}.

The rest of the paper is organized as follows.
Section \ref{prover} overviews our Prolog-based theorem prover and the sequent calculus it is derived from.
Section \ref{abdu} introduces our {\em protasis synthesizer} and its uses for abductive reasoning.
Section \ref{mincan} generalizes our approach to the synthesis of minimal canonical assumptions.
Section \ref{disc} discusses significance of our results, its possible extensions as well as some of its limitations.
Section \ref{rel} overviews related work and section \ref{concl} concludes the paper.

We assume the reader is fluent in Prolog, propositional intuitionistic and classical logic and familiar with abductive reasoning and key concepts behind sequent calculi and automated theorem proving.

\section{Background: The Intuitionistic Propositional Logic Theorem Prover}\label{prover}
We will derive our Prolog prover from a set of compact and elegant sequent calculus rules formally describing provability in {\bf IL}. 

\subsection{Roy Dyckhoff's {\bf G4ip} calculus}

Motivated by problems related to loop avoidance in implementing  Gentzen's {\bf LJ} calculus,
Roy Dyckhoff designed and proved sound and complete a sequent calculus-based axiomatization of {\bf IL} \cite{dy1} . He has  proved that the calculus is terminating, by identifying a multiset
ordering-based formula size definition that decreases after each step \cite{dy1}.

The sequents of the {\bf G4ip} calculus follow:
\[
 \begin{array}{lll}
 \deduce{\Ga,p \seq p}{} \ \ \ \text{{\it Ax} \ \ ($p$ an atom)}  & 
  \deduce{\Ga,\bot\seq \De}{} \ \ \ L\bot \\
 \\ 
 \infer[R\en]{\Ga \seq \phi \en \psi}{\Ga\seq \phi & \Ga \seq \psi} & 
  \infer[L\en]{\Ga, \phi\en \psi \seq \De}{\Ga, \phi, \psi \seq \De} \\
 \\
 \infer[R\!\of \ (i=0,1)]{\Ga \seq \phi_0 \of \phi_1}{\Ga \seq \phi_i} & 
  \infer[L\of]{\Ga,\phi\of \psi\seq \De}{\Ga, \phi \seq \De & \Ga,\psi \seq \De} \\
  \\
 \infer[R\!\imp]{\Ga \seq \phi \imp \psi}{\Ga,\phi \seq \psi} & 
 \infer[Lp\!\imp\text{ ($p$ an atom)}]{\Ga, p,p \imp \phi \seq \De}{\Ga,p,\phi \seq \De}\\
 \\
 \infer[L\en\!\imp]{\Ga, \phi\en\psi \imp \gamma \seq \De}{\Ga,\phi\imp (\psi\imp\gamma)\seq\De} & 
 \infer[L\of\!\imp]{\Ga,\phi \of \psi \imp \gamma \seq \De}{
  \Ga,\phi \imp \gamma, \psi \imp \gamma \seq \De}\\ 
 \\ 
 \infer[L\!\imp\!\imp]{\Ga, (\phi\imp \psi) \imp \gamma \seq \De}{
  \Ga, \psi\imp \gamma \seq \phi \imp \psi & \gamma,\Ga \seq \De}\\
\end{array}
\] 
Key to the termination proof in \cite{dy1} is the rule $L\!\imp\!\imp$ that breaks down nested implications into ``smaller'' ones, each containing fewer connectives.
The rules work with the context $\Gamma$
being a multiset,  but it has been shown later \cite{dy2} that $\Gamma$ can be  a set, with duplication in contexts 
eliminated.

Note that the same calculus has been discovered independently in the 50's by Vorob'ev and
in the 80's-90's by Hudelmaier \cite{hud88}. 

\subsection{Implementing the Theorem Prover}\label{deriv}

In the tradition of "lean theorem provers", we can build one directly
from the {\bf G4ip} calculus, in a goal oriented style, by reading the rules {\em from conclusions to premises}.

Thus, we start with a simple, almost literal translation of sequent rules to Prolog with values in the environment $\Gamma$ denoted by the variable {\tt Vs}. Besides implication (denoted \verb~->~), conjunction (denoted {\verb~&~}) and disjunction (denoted \verb~v~), we  implement rules for inverse implication (denoted \verb~<-~), negation (denoted \verb|~|) and equivalence (denoted \verb~<->~). We also add rules for a top element (denoted \verb~true~) and a bottom element (denoted \verb~false~). 

Correctness of our additions follows from the definitions of:

\BI
\I intuitionistic inverse implication:
$\phi \leftarrow \psi ~\equiv~ \psi \rightarrow \phi$
\I intuitionistic equivalence: 
$\phi \leftrightarrow \psi \equiv \phi \rightarrow \psi ~ \verb~&~ ~  \psi \rightarrow \phi$
\I intuitionistic negation: $ \verb|~|\phi ~\equiv~ \phi \rightarrow \verb~false~$.
\I top element: $\verb~true~ ~\equiv~ \verb~false~ \rightarrow  \verb~false~ $.
\EI

Note that the added connectives are meant to enhance the expressiveness of the logic. For instance, ``\verb~<->~'' allows expressing the fact that two formulas are equivalent and thus equiprovable, ``\verb~head <- body~'' mimics Prolog's familiar Horn clause syntax ``\verb~head :- body~'' and finally the negation symbol 
makes formulas more compact and human-readable.

We define operators for all our connectives, except \verb~->~, that we will use with its standard right associativity.

\begin{code}
:- op(525,  fy,  ~ ).
:- op(550, xfy,  & ).    
:- op(575, xfy,  v ).    
:- op(600, xfx,  <-> ).  
:- op(800, yfx,  <- ).   
\end{code}
A formula {\tt T} is a theorem if it is provable from an empty set of assumptions:
\begin{code}
iprover(T) :- iprover(T,[]).
\end{code}
We follow here Dyckhoff's calculus but delegate details to helper predicates
{\tt iprover\_reduce/4} and {\tt iprover\_impl/4}. 
\begin{code}
iprover(true,_):-!.
iprover(A,Vs):-memberchk(A,Vs),!.
iprover(_,Vs):-memberchk(false,Vs),!.
iprover(~A,Vs):-!,iprover(false,[A|Vs]).
iprover(A<->B,Vs):-!,iprover(B,[A|Vs]),iprover(A,[B|Vs]).
iprover((A->B),Vs):-!,iprover(B,[A|Vs]).
iprover((B<-A),Vs):-!,iprover(B,[A|Vs]).
iprover(A & B,Vs):-!,iprover(A,Vs),iprover(B,Vs).
iprover(G,Vs1):- 
  select(Red,Vs1,Vs2),
  iprover_reduce(Red,G,Vs2,Vs3),
  !,
  iprover(G,Vs3).
iprover(A v B, Vs):-(iprover(A,Vs) ; iprover(B,Vs)),!.
\end{code}
{\tt iprover\_reduce/4} is the first step in breaking down formulas in the premise into
their components.
\begin{code}
iprover_reduce(true,_,Vs1,Vs2):-!,iprover_impl(false,false,Vs1,Vs2).
iprover_reduce(~A,_,Vs1,Vs2):-!,iprover_impl(A,false,Vs1,Vs2).
iprover_reduce((A->B),_,Vs1,Vs2):-!,iprover_impl(A,B,Vs1,Vs2).
iprover_reduce((B<-A),_,Vs1,Vs2):-!,iprover_impl(A,B,Vs1,Vs2).
iprover_reduce((A & B),_,Vs,[A,B|Vs]):-!.
iprover_reduce((A<->B),_,Vs,[(A->B),(B->A)|Vs]):-!.
iprover_reduce((A v B),G,Vs,[B|Vs]):-iprover(G,[A|Vs]).
\end{code}
{\tt iprover\_impl/4} details the  case analysis of the
handling of implication (and its instances) prescribed by rule
$L\!\imp\!\imp$.
\begin{code}
iprover_impl(true,B,Vs,[B|Vs]):-!.
iprover_impl(~C,B,Vs,[B|Vs]):-!,iprover((C->false),Vs).
iprover_impl((C->D),B,Vs,[B|Vs]):-!,iprover((C->D),[(D->B)|Vs]).
iprover_impl((D<-C),B,Vs,[B|Vs]):-!,iprover((C->D),[(D->B)|Vs]).
iprover_impl((C & D),B,Vs,[(C->(D->B))|Vs]):-!.
iprover_impl((C v D),B,Vs,[(C->B),(D->B)|Vs]):-!.
iprover_impl((C<->D),B,Vs,[((C->D)->((D->C)->B))|Vs]):-!.
iprover_impl(A,B,Vs,[B|Vs]):-memberchk(A,Vs).    
\end{code}

\BX After defining:
\begin{code}
iprover_test:-
   Taut = ((p & q) <-> (((p v q)<->q)<->p)), iprover(Taut),
   Contr=(a & ~a), \+ (iprover(Contr)).
\end{code}
we observe success on proving a tautology and failing to prove a contradiction, 
but we refer to \cite{padl19} for an extensive combinatorial testing of a variant of
this prover as well as its testing against the ILTP benchmark\footnote{\url{http://www.iltp.de/}} and several other provers.
\EX

\subsection{Classical Logic For Free}
Glivenko's theorem states that {\em a propositional formula $F$ is a classical tautology if and only if $\tilde ~~ \tilde ~ F$ is an intuitionistic tautology}. This gives us a classical prover for free, that we will use as an alternative to {\tt iprove} when defining several concepts parameterized by a prover.
\begin{code}
cprover(T):-iprover( ~ ~T).
\end{code}

\BX The two provers, as it is well known, will disagree on \verb|p v ~p|
\begin{codex}
?- iprover(p v ~p).
false.
?- cprover(p v ~p).
true.
\end{codex}
but agree on \verb|p & ~p|:
\begin{codex}
?-iprover(p & ~p).
false.
?-cprover(p & ~p).
false.
\end{codex}
\EX

\section{Abductive Reasoning Mechanisms}\label{abdu}
We are now ready to introduce our search for assumptions that make a given formula an intuitionistic tautology.

\subsection{Generating the Abducibles}
Defining some of the atoms occurring in a formula $F$ as the only ones to be used in the search process brings us to declare them as {\em abducibles} \cite{eshghi89}, but we will also enable the option to make abducible all the atoms occurring in $F$.
Thus, when the variable {\tt Abducibles} is free, 
all atomic symbols will be considered abducibles.

\begin{code}
abducibles_of(Formula,Abducibles):-var(Abducibles),!,atoms_of(Formula,Abducibles).
abducibles_of(_,_).
\end{code}
The predicate {\tt atoms\_of/2} finds the set of all the atoms occurring in a formula. It backtracks over all arguments, recursively and it collects atomic elements to a list, with {\tt setof/3} which also eliminates possible duplicates.
\begin{code}
atom_of(A,R):-atomic(A),!,R=A.
atom_of(T,A):-arg(_,T,X),atom_of(X,A).

atoms_of(T,As):-setof(A,atom_of(T,A),As).
\end{code}

\subsection{Protasis Generation}

If $F$ is a formula, we can think of a premise $C$  such that $C \rightarrow F$ is a theorem\footnote{or, equivalently, when $C$ is the {\em premise} and $F$ is the {\em conclusion} of a provable sequent}  as a counterfactual assumption making $F$ conditionally true. We call such a formula $C$ a {\em protasis}. Thus, given a formula and a set of assumptions (our abducibles), we will need to search among them for assumptions  that would make the formula a theorem.

The predicate {\tt any\_protasis/6} implements this idea, subject to a set of parameters:
\BI
\I {\tt Prover} allows a choice of the underlying logic (e.g., intuitionistic or classical)
\I {\tt AggregatorOp} fixes one of the connectives of the logic, from which the protasis is built
\footnote{the restriction to on operator will be lifted later in section\ref{mincan}, when we generalize this mechanism to a set of canonical formulas}
\I The {\tt yes/no} flag {\tt WithNeg} decides if negations of the abducibles can be part of the protasis
\I {\tt Abducibles} is a set of atoms or a free variable, meaning that all atoms will be included
\I {\tt Assumption} will be any protasis, that given the previously specified parameters, ensures that {\tt Assumption->Formula} is a theorem in the logic specified by {\tt Prover}.
\EI
\begin{code}
any_protasis(Prover,AggregatorOp,WithNeg,Abducibles,Formula,Assumption):-
  abducibles_of(Formula,Abducibles),
  mark_hypos(WithNeg,Abducibles,Literals),
  subset_of(Literals,Hypos),
  join_with(AggregatorOp,Hypos,Assumption),
  \+ (call(Prover,Assumption->false)), 
  call(Prover,Assumption->Formula).    
\end{code}
The predicate {\tt mark\_hypos/3} will mark with their negations the abducibles if we want to allow them to occur in the protasis positively or prefixed by their negations.
\begin{code}
mark_hypos(_,[],[]).
mark_hypos(yes,[P|Ps],[P,~P|Ns]):-mark_hypos(yes,Ps,Ns).
mark_hypos(no,[P|Ps],[P|Ns]):-mark_hypos(no,Ps,Ns).
\end{code}
Our subset generator {\tt subset\_of}, used to iterate over all subsets of the abducibles, first enumerates templates of increasing length as we want smaller subsets to be tried first.
\begin{code}
subset_of(Xs,Ts):-template_from(Xs,Ts),tsubset(Xs,Ts).

template_from(_,[]).
template_from([_|Xs],[_|Zs]):-template_from(Xs,Zs).
\end{code}
Then, for each template of length $K$, it fills it with a subset of length $K$ of the $N$ abducibles, one at a time, on backtracking.
\begin{code}
tsubset([],[]).
tsubset([X|Xs],[X|Rs]):-tsubset(Xs,Rs).
tsubset([_|Xs],Rs):-tsubset(Xs,Rs).
\end{code}
The predicate {\tt join\_with\_op} builds an expression from a sequence of abducible literals (atoms, possibly negated) with a given operator.
\begin{code}
join_with_op(_,[],true).
join_with_op(_,[X],X).
join_with_op(Op,[X,Y|Xs],R):-join_with_op(Op,[Y|Xs],R0),R=..[Op,X,R0].
\end{code}
How we join the abducible literals with help of a given operator, is different for associative and commutative operators like {\tt \&} and {\tt v} (the default 3-rd clause of {\tt join\_with}) and  {\tt ->}, {\tt <-}, {\tt <->}, that we treat as special cases.

Thus, once the head was picked with {\tt select/3}, the order of the remaining literals is immaterial.
\begin{code}
join_with(Op,Xs,R):-
  memberchk(Op,[(->),(<-)]),!,
  select(Head,Xs,Ys), append(Ys,[Head],Zs),
  join_with_op((->),Zs,R).
\end{code}
For non-associative {\tt <->} we will use all permutations of the abducibles.
\begin{code}
join_with(Op,Xs,R):-Op=(<->),!,permutation(Xs,Ys),join_with_op(Op,Ys,R).
\end{code}
Finally, as {\tt v} and {\tt \&} are permutation invariant, we just call {\tt join\_with\_op}.
\begin{code}
join_with(Op,Xs,R):- join_with_op(Op,Xs,R).
\end{code}
Implication and reverse implications are handled in {\tt join\_with}, knowing that their components, except the head, are permutation invariant, with the following equivalences in mind:
\begin{align*}
\phi_0 \leftarrow \phi_1 \dots  \leftarrow \phi_n \equiv \phi_0 \leftarrow \phi_1 ~\&~ \dots ~\&~ \phi_n
\end{align*} 
and
\begin{align*}
\phi_n \rightarrow \phi_{n-1} \dots  \rightarrow \phi_1 \rightarrow \phi_0 ~\equiv~ \phi_n ~\&~ \phi_{n-1} ~\&~ \dots ~\&~ \phi_1 ~\rightarrow \phi_0
\end{align*}
Note that these hold both intuitionistically and classically, as it can be quickly verified with {\tt iprover} and {\tt cprover}.

\subsection{The Weakest Protasis}

The next step is defining a {\em weakest protasis}, keeping in mind that a {\em partial order}\footnote{in contrast to classical logic, where {\tt (p->q) v (q->p)} is a theorem} among them is defined by the intuitionistic implication ({\tt->}).
First, we will collect with {\tt setof/3} all the candidate assumptions to ensure that the prover is called on each only once.
\begin{code}
weakest_protasis(Prover,AggregatorOp,WithNeg,Abducibles,Formula,Assumption):-
  setof(Assumption,
    any_protasis(Prover,AggregatorOp,WithNeg,Abducibles,Formula,Assumption),
    Assumptions),
  weakest_with(Prover,Assumptions,Assumption).
\end{code}
Next, we ensure that a {\em weakest protasis} is such that it does not imply any other protasis, thus that it is a minimal element w.r.t. our partial order. We rely for that on the predicate {\tt weakest\_with/3}:
\begin{code}
weakest_with(_,Gs,G):-memberchk(true,Gs),!,G=true.
weakest_with(Prover,Gs,G):-
   select(G,Gs,Others),
   \+ (member(Other,Others),
       weaker_with(Prover,Other,G)).
\end{code}
The partial order relation is exposed as the predicate {\tt weaker\_with/3} which ensures  that a weakest protasis does not imply any other protasis, including ones that might imply it.
\begin{code}
weaker_with(Prover,P,Q):- \+ call(Prover,(P->Q)), call(Prover,(Q->P)).
\end{code}
Note  that the predicate {\tt weakest\_protasis/6} depends on the same parameters as {\tt any\_protasis}, in particular on the {\tt Prover} implementing a given provability relation.
\BX
Peirce's law is known to hold in {\bf CL} and not hold in {\bf IL}. In fact, it can turn {\bf IL} into {\bf CL} if added as an axiom. The predicate {\tt peirce/2} will try to synthesize an assumption that would make it hold.
\begin{code}
peirce(Prover,WhatIf):-
    Formula=(((p->q)->p)->p),
    WithNeg=yes, AggregatorOp=(v), Abducibles=[p],
    weakest_protasis(Prover,AggregatorOp,WithNeg,Abducibles,Formula,WhatIf).
\end{code}
When running it we get:
\begin{codex}
?- peirce(iprover,Protasis).
Protasis = p v ~p.
?- peirce(cprover,Protasis).
Protasis = true.
\end{codex}
This reveals an interesting fact. It tells us that if \verb|p v ~p| were assumed, Peirce's law would hold in {\bf IL}, as {\tt iprove} would succeed. As it is well known, \verb|p v ~p| would turn {\bf IL} into {\bf CL} and {\tt cprove} tells us that indeed, Peirce's law holds unconditionally in {\bf CL}.

\BX
We can also synthesize conditional assumptions when using implication or inverse implication as our aggregator connective. After defining:
\begin{code}
impl_aggr(H):-
   T=(a<-((a<-(b<-d))&(b<-c))),
   Prover=iprover, WithNeg=yes, AggregatorOp=(->), As=[c,d],
   weakest_protasis(Prover,AggregatorOp,WithNeg,As,T,H).
\end{code}
we obtain:
\begin{codex}
?- impl_aggr(H).
H =  (d->c).
\end{codex}
showing that the implication {\tt d->c} (equivalent of the Horn Clause {\tt c:-d}) should hold for the formula to be a theorem. This illustrates a mechanism to synthesize Horn Clauses playing the role of conditional assumptions.
\EX
Another interesting case is that of a contradiction. After defining {\tt contra\_test/1} as:
\begin{code}
contra_test(H):-
   T = (p & ~p),
   Prover=iprover, WithNeg=yes, AggregatorOp=(&),
   weakest_protasis(Prover,AggregatorOp,WithNeg,_Abducibles,T,H).
\end{code}
and running it with:
\begin{codex}
?- contra_test(Protasis).
false.
\end{codex}
\EX
we can see that no assumption would make the contradiction a tautology, and this will also be the case for {\tt Prover=cprover}. Note that to ensure this, we have enforced in the definition of {\tt any\_protasis} that such assumptions should themselves not be contradictions.

\BX
In the case of the logic of here-and-there \cite{pearce97}, derived from {\bf IL} by adding the axiom\\
\verb| f v (f->g) v ~g|\\
\noindent as shown in \cite{lifschitz01},  we will get with {\tt cprover} the protasis {\tt true}. This indicates, as expected, that it is already a theorem in {\bf CL} and thus also a theorem in the  logic of here-and-there.
\begin{codex}
?- weakest_protasis(cprover,(v),yes,_,(f v (f->g) v ~g),P).
P = true.
\end{codex}
On the other hand, the less obvious weakest protasis obtained for {\tt iprove} indicates that the excluded middle rule would be needed for both {\tt f} and {\tt g}.
\begin{codex}
?- weakest_protasis(iprover,(v),yes,_,(f v (f->g) v ~g),P).
P = f v ~f v g v ~g.
\end{codex}
\EX

\subsection{An Example of Intuitionistic Abductive Reasoning}

With the logic of synthesizing meaningful minimal assumptions that make a formula a theorem clarified, we are now ready to revisit  abductive reasoning along the lines of
\cite{eshghi89}. 

The predicate {\tt explain\_with/5} finds, given a {\tt Prover},  the abductive inference problem parameterized by:
\BI
\I a formula  {\tt Prog} seen here as representing a knowledge base
\I a set of {\tt Abducibles} occurring in {\tt Prog}
\I a goal formula {\tt G} such that {\tt Prog -> G} should always hold
\I a formula {\tt IC}  playing the role of integrity constraints meant to filter out unwanted assumptions
\EI
\begin{code}
explain_with(Prover,Abducibles,Prog,IC,G,Expl):-
    any_protasis(Prover,(&),yes,Abducibles,(Prog->G), Expl),
    call(Prover, Expl & Prog->G),
    call(Prover,(Expl & Prog->IC)),
    \+ (call(Prover,(Expl & Prog -> false))).
\end{code}
Note also that {\tt any\_protasis}  replaces {\tt weakest\_protasis} in this definition, given that minimality might want to be stated as part of the integrity constraints {\tt IC}.
\BX
To revisit a simple example of abductive explanation generation, we define:

\begin{code}
why_wet(Prover):-
    IC = ~(rained & sunny),
    P = sunny & (rained v sprinkler -> wet), As=[sprinkler,rained], G = wet,
    writeln(prog=P), writeln(ic=IC),
    explain_with(Prover,As,P,IC,G,Explanation),
    writeln('Explanation:' --> Explanation).
\end{code}
\noindent Then, when running it, it will display, as expected:
\begin{codex}
?- why_wet(iprover).
prog=sunny&(rained v sprinkler->wet)
ic= ~ (rained&sunny)
Explanation: --> sprinkler& ~rained\end{codex}
\EX

\section{Synthesis of Minimal Canonical Assumptions}\label{mincan}

We will now generalize our abductive reasoning mechanism by lifting the constraint on the premise of our sequent from literals connected by a single operation to a canonical form that has been shown to be able to represent arbitrary {\bf IL} formulas.

\subsection{The Mints Transformation}

Grigori Mints has proven, in his seminal paper studying complexity classes for 
intuitionistic propositional logic \cite{mints92}, that any formula $f$
is equiprovable to a formula of the form $X_f \rightarrow g$, where
$X_f$ is a conjunction of formulas of one of the forms:
\begin{align*}
p, \char`\~p,~ p \rightarrow q,~(p \rightarrow q)  \rightarrow r,~ p \rightarrow (q \rightarrow r),~  p \rightarrow (q ~v~ r),~ p \rightarrow  \char`\~q,~  \char`\~q \rightarrow p. 
\end{align*}
\noindent With introduction of new variables (like with the Tseitin transform for SAT or ASP solvers), the transformation runs in linear space and time.
Note that as a premise to a sequent can be seen as a conjunction implying its conclusion, the conjunction of the formulas described by Mints can be seen as serving the same purpose as the conjunctive normal form ({\bf CNF}) in classical logic.

Thus, by generating this set of bounded size formulas as premises of a sequent, we can express equivalent formulas of unbounded size, otherwise subject to a much larger search space in the formula synthesis process.  

We will now generate, using a given set of abducibles, premises built of these formulas, with their propositional variables selected from the set of abducibles.

Conceptually, while we will keep calling the propositional variables involved in our premise {\em abducibles}, we shall see them from now on simply as a set of {\em independent variables} on which the  derivation of the sequent's conclusion from its premise depends.

We define a generator for the set of {\em Mints formulas} with help of a DCG grammar\footnote{
As a note to the reader unfamiliar with Prolog's Definite Clause Grammars (DCG) preprocessor,
it transforms a DCG clause like \verb~a --> b,c,d~ into an ordinary Prolog clause
\verb~a(S0,Sn) :- b(S0,S1),c(S1,S2),d(S2,Sn)~, to conveniently keep track of state changes in the ``chained'' variables \verb~S0,S1,...,Sn~.}
that
collects its propositional variables from a list of abducibles. 

\begin{code}
mints_formula(P)-->[P].                    mints_formula(~P)-->[P].
mints_formula((P->Q))-->[P,Q].             mints_formula((P->Q)->R)-->[P,Q,R].
mints_formula((P->(Q->R)))-->[P,Q,R].      mints_formula((P->(Q v R))) -->[P,Q,R].
mints_formula((P-> ~Q))-->[P,Q].           mints_formula((~P->Q))-->[P,Q].
\end{code}

We extend our DCG, subject to the same limitation on available abducibles, to generate a list of formulas meant to be part of the premise. We will later aggregate this list into a conjunction.
\begin{code}
mints_conjuncts([])-->[].
mints_conjuncts([F|Fs])-->mints_formula(F),mints_conjuncts(Fs).
\end{code}
Next, we eliminate duplicates with Prolog's {\tt sort/2}.
\begin{code}
mints_conjuncts(Atoms,Conjuncts):-mints_conjuncts(Ps,Atoms,[]),sort(Ps,Conjuncts).
\end{code}
We derive {\tt any\_mints\_premise/4} in a way similar to {\tt any\_protasis/6}, except that the arguments {\tt WithNeg} and {\tt AggregatorOp} become unnecessary and multiple occurrences of any abducible need to be supported. For brevity, we explain our steps as comments in the code.
\begin{code}
any_mints_premise(Prover,Abducibles,Formula,Premise):-
  abducibles_of(Formula,Abducibles),
  subset_of(Abducibles,Chosen),     
  template_from(Abducibles,Atoms),  
  part_as_equiv(Atoms,Chosen),      
  mints_conjuncts(Atoms,Conjuncts), 
  join_with_op((&),Conjuncts,Premise), 
  \+ (call(Prover,Premise->false)),    
  call(Prover,Premise->Formula).       
\end{code}
Note that we have limited the length of the premise in the case of the Mints-formulas, arbitrarily, to the number of abducible atoms, that the selection of Atoms has as an upper bound. For more flexibility, this can be lifted to be based on a length parameter passed to {\tt any\_mints\_premise}.

\subsection{Labeling the Variables in the Mints Formulas}

We will describe here the implementation of {\tt part\_as\_equiv/2} that will be used to
to generate variables bound to possibly repeated occurrences of each atom.
Note that equalities of logic variables define equivalence 
classes that  correspond to partitions
of the set of variables. We implement this simply by
selectively unifying them. 

The predicate {\tt part\_as\_equiv/2}
takes a list of distinct logic variables
and generates partitions-as-equivalence-relations
by unifying them ``nondeterministically''.
It also collects the unique variables defining
the equivalence classes, as a list given by its second argument. It  works reversibly,
when unique values are given as its second argument and a bound list of free variables as its first.
\begin{code}
part_as_equiv([],[]).
part_as_equiv([U|Xs],[U|Us]):-complement_of(U,Xs,Rs),part_as_equiv(Rs,Us).
\end{code}
To implement it,
we  split a set repeatedly in subset+complement
pairs with help from the predicate {\tt complement\_of/2}.
\begin{code}
complement_of(_,[],[]).
complement_of(U,[X|Xs],NewZs):-complement_of(U,Xs,Zs),place_element(U,X,Zs,NewZs).

place_element(U,U,Zs,Zs).
place_element(_,X,Zs,[X|Zs]).
\end{code}

\BX
Here, we are interested in the reverse use of {\tt part\_as\_equiv}, 
with the list of unique variables
as input and a sequence of variables of fixed length but possibly repeated
occurrences as output.
\begin{codex}
?- length(Vs,4),part_as_equiv(Vs,[a,b]).
Vs = [a, b, a, a] ; Vs = [a, a, b, a] ; Vs = [a, b, b, a] ;
Vs = [a, a, a, b] ; Vs = [a, b, a, b] ; Vs = [a, a, b, b] ; Vs = [a, b, b, b] .
\end{codex}
\EX

We derive {\tt weakest\_mints\_premise/4} in a way similar to {\tt weakest\_protasis/6} except for passing only the relevant arguments to {\tt any\_mints\_premise/4}.
\begin{code}
weakest_mints_premise(Prover,Abducibles,Formula,Premise):-
  setof(Premise,
    any_mints_premise(Prover,Abducibles,Formula,Premise),
    Premises),
  weakest_with(Prover,Premises,Premise).
\end{code}

\BX
When using the axiom that conservatively extends {\bf IL} to the logic of here-and-there \cite{lifschitz01} we observe again that no premise is needed for {\tt cprover} and that {\tt iprover} suggests as premises either one of the disjuncts in the axiom, or, more interestingly, \verb|g-> ~g|. 
\begin{codex}
?- weakest_mints_premise(cprover,_,(f v (f->g) v ~g),P).
P = true.
?- weakest_mints_premise(iprover,_,(f v (f->g) v ~g),P).
P = f ; P = ~g ; P =  (f->g) ; P =  (g-> ~g).
\end{codex}
\EX
In fact, we observe:
\begin{codex}
?- iprover((g -> ~g) <-> ~g).
true.
\end{codex}
suggesting   that extending {\bf IL} with:\\
\verb|f v (f->g) v (g -> ~g)|\\
would result in an alternative axiomatization of the logic of here-and-there.

\section{Discussion}\label{disc}
We hope that the astute reader is aware at this point that the paper is an exploration of the theory behind some fundamental concepts relating abductive reasoning, program synthesis and theorem proving in a concise and easily replicable form, facilitated by the choice of Prolog as our meta-language, but with the possibility of a fairly routine transliteration to a traditional ``formulas-on-paper'' presentation in mind.
As such, the paper can be seen as an executable specification of these concepts. While not neglecting  minimal efforts for efficient execution, and some elegance in the coding style, our main priority was to ensure that the paper conveys its message as a fully self-contained literate program.

We have designed our abductive reasoning logic entirely in a proof-theoretical framework, in contrast to the  usual model-theoretical semantics, arguably in the original spirit of intuitionistic logic.

Both our weakest protasis-based and weakest Mints formulas-based synthetic assumptions are attempts to recover in {\bf IL} an analogue of the {\bf CNF} available for a formula in {\bf CL}. At the same time,  finding the weakest assumptions shares the focus on minimal models encountered in various logic calculi. Our interest in finding weakest assumptions under which the formula becomes a theorem is driven by the transitivity of the partial order induced by {\tt ->}, given that for a given formula {\tt f}, becoming a theorem under the assumption {\tt w}, ensures also that if {\tt (s->w)}, then {\tt (s->f)} is also a theorem, where {\tt w} denotes a weakest assumption and {\tt s} denotes a (stronger) assumption that implies it.

We have restricted ourselves to propositional logic but we foresee extensions to stronger  logics among which Monadic First Order Logic (known as decidable for {\bf CL} and undecidable for {\bf IL}), enhanced with Prolog's constraint solving mechanisms is a promising option.

While we have forced a clear separation between our meta-language (Prolog) and object-language ({\bf IL}), it would be quite easy to extend our theorem prover to reflect Prolog's negation as failure in the object-language as an addition to {\bf IL}. This would result in a logic with two flavors of negation, similar in the context of {\bf IL} to the underlying equilibrium logic of {\bf ASP}.

\section{Related work}\label{rel}


We refer to \cite{abd_lp} as  still the most lucid and comprehensive overview (also citing 124 papers) on abductive reasoning in Logic Programming and to \cite{eshghi89} as one of the most influential initiators for the interest in the field, with connections explored in depth to negation as failure and non-monotonic reasoning. Some of our examples related to the logic of here-and-there and equilibrium logic \cite{pearce96,table5valued} originate in \cite{lifschitz01}, where intermediate logics relevant for the foundation of {\bf ASP} systems are overviewed. For abductive reasoning in the context of several logics including non-monotonic ones, we mention  \cite{gabbay2002goal} and \cite{gabbay2005goal}.

By contrast, the  novelty of our approach is the generalized view of abductive reasoning as an instance of program synthesis controlled by a theorem prover. Our theorem prover is derived directly from the {\bf G4ip} sequent calculus \cite{dy1,dy2}. In \cite{padl19} details of this derivation process as well as a combinatorial testing framework used to insure  correctness are given. The idea of using a theorem-prover for the synthesis of modal formulas is also present in \cite{lopstr20}, having as an outcome an embedding of the  epistemic logic {\bf IEL} in {\bf IL} and a derived theorem prover for that logic. In \cite{iclp20} a theorem prover, restricted to the implicational fragment of {\bf IL} is used to derive a theorem prover for implicational linear logic, with help from the Curry-Howard correspondence and the use of linearity of the resulting lambda terms as a filtering mechanism. In this context, the distinct focus of the current paper is on a very general formula synthesis mechanism within {\bf IL} itself, covering abductive reasoning  and emulating in {\bf IL} key semantic concepts available in  {\bf CL} and intermediate logics.

\section{Conclusions}\label{concl}
We have presented a fully executable specification of a generalized abductive reasoning framework, that can be relatively easily ported to any logic  for which a decision mechanism exists (e.g., as provided by a theorem prover). In particular, this applies to several interesting intermediate logics among which the equilibrium-logic (relevant as a foundation  of {\bf ASP} systems) as well as modal logics and their instantiations as alethic, temporal, deontic or epistemic systems. Besides providing (in the form of the concept of weakest protasis) an analogue of the unavailable truth-table models for intuitionistic formulas, we have also generalized our abduced sequent premises to use minimal canonical formulas to which arbitrary {\bf IL} formulas can be broken down, with the potential of synthesizing  salient assumptions that would make a given formula a theorem. When the underlying logic is  used to model a set of safety constraints that should always hold, this generalized abduction synthesis could reveal critical missing assumptions, not just as literals but also as a conjunction of interdependencies among them.

\section*{Acknowledgement} 
We thank the anonymous reviewers of ICLP'2022 for their constructive comments and suggestions.

\bibliographystyle{acmtrans}
\bibliography{tarau,theory,proglang,biblio}

\begin{thebibliography}{}

\bibitem[\protect\citeauthoryear{Denecker and Kakas}{Denecker and
  Kakas}{2002}]{abd_lp}
{\sc Denecker, M.} {\sc and} {\sc Kakas, A.} 2002.
\newblock Abduction in logic programming.
\newblock In {\em {Computational Logic: Logic Programming and Beyond}}.
  Springer, 402--36.

\bibitem[\protect\citeauthoryear{Dyckhoff}{Dyckhoff}{1992}]{dy1}
{\sc Dyckhoff, R.} 1992.
\newblock {Contraction-free sequent calculi for intuitionistic logic}.
\newblock {\em Journal of Symbolic Logic\/}~{\em 57,\/}~3, 795--807.

\bibitem[\protect\citeauthoryear{Dyckhoff}{Dyckhoff}{2016}]{dy2}
{\sc Dyckhoff, R.} 2016.
\newblock {Intuitionistic Decision Procedures Since Gentzen}.
\newblock In {\em Advances in Proof Theory}, {R.~Kahle}, {T.~Strahm}, {and}
  {T.~Studer}, Eds. Springer International Publishing, Cham, 245--267.

\bibitem[\protect\citeauthoryear{Eshghi and Kowalski}{Eshghi and
  Kowalski}{1989}]{eshghi89}
{\sc Eshghi, K.} {\sc and} {\sc Kowalski, R.~A.} 1989.
\newblock {Abduction Compared with Negation by Failure}.
\newblock In {\em Logic Programming, Proceedings of the Sixth International
  Conference, Lisbon, Portugal, June 19-23, 1989}, {G.~Levi} {and}
  {M.~Martelli}, Eds. {MIT} Press, 234--254.

\bibitem[\protect\citeauthoryear{Gabbay and Olivetti}{Gabbay and
  Olivetti}{2002}]{gabbay2002goal}
{\sc Gabbay, D.} {\sc and} {\sc Olivetti, N.} 2002.
\newblock Goal-oriented deductions.
\newblock In {\em Handbook of Philosophical Logic}. Springer, 199--285.

\bibitem[\protect\citeauthoryear{Gabbay}{Gabbay}{2000}]{gabbay2005goal}
{\sc Gabbay, D.~M.} 2000.
\newblock {Goal Directed Mechanisms: Proofs, Interpolation and Abduction
  Procedures}.
\newblock In {\em Proceedings of the Seventh Workshop on Automated Reasoning,
  Bridging the Gap between Theory and Practice, King's College London, UK,
  20-21 July 2000}, {H.~J. Ohlbach}, {U.~Endriss}, {O.~Rodrigues}, {and}
  {S.~Schlobach}, Eds. {CEUR} Workshop Proceedings, vol.~32. CEUR-WS.org.

\bibitem[\protect\citeauthoryear{Hudelmaier}{Hudelmaier}{1988}]{hud88}
{\sc Hudelmaier, J.} 1988.
\newblock {\em A PROLOG Program for Intuitionistic Logic}.
\newblock SNS-Bericht-. Universit{\"a}t T{\"u}bingen.

\bibitem[\protect\citeauthoryear{Lifschitz, Pearce, and Valverde}{Lifschitz
  et~al\mbox{.}}{2001}]{lifschitz01}
{\sc Lifschitz, V.}, {\sc Pearce, D.}, {\sc and} {\sc Valverde, A.} 2001.
\newblock Strongly equivalent logic programs.
\newblock {\em {ACM} Trans. Comput. Log.\/}~{\em 2,\/}~4, 526--541.

\bibitem[\protect\citeauthoryear{Mints}{Mints}{1992}]{mints92}
{\sc Mints, G.} 1992.
\newblock {Complexity of Subclasses of the Intuitionistic Propositional
  Calculus}.
\newblock {\em {BIT}\/}~{\em 32,\/}~1, 64--69.

\bibitem[\protect\citeauthoryear{Pearce}{Pearce}{1996}]{pearce96}
{\sc Pearce, D.} 1996.
\newblock A new logical characterisation of stable models and answer sets.
\newblock In {\em {NMELP}}. Lecture Notes in Computer Science, vol. 1216.
  Springer, 57--70.

\bibitem[\protect\citeauthoryear{Pearce}{Pearce}{1997}]{pearce97}
{\sc Pearce, D.} 1997.
\newblock A new logical characterisation of stable models and answer sets.
\newblock In {\em Selected Papers from the Non-Monotonic Extensions of Logic
  Programming}. NMELP '96. Springer-Verlag, Berlin, Heidelberg, 57--70.

\bibitem[\protect\citeauthoryear{Pearce, de~Guzm{\'{a}}n, and Valverde}{Pearce
  et~al\mbox{.}}{2000}]{table5valued}
{\sc Pearce, D.}, {\sc de~Guzm{\'{a}}n, I.~P.}, {\sc and} {\sc Valverde, A.}
  2000.
\newblock { Tableau Calculus for Equilibrium Entailment}.
\newblock In {\em Automated Reasoning with Analytic Tableaux and Related
  Methods, International Conference, {TABLEAUX} 2000, St Andrews, Scotland, UK,
  July 3-7, 2000, Proceedings}, {R.~Dyckhoff}, Ed. Lecture Notes in Computer
  Science, vol. 1847. Springer, 352--367.

\bibitem[\protect\citeauthoryear{Tarau}{Tarau}{2019}]{padl19}
{\sc Tarau, P.} 2019.
\newblock {A Combinatorial Testing Framework for Intuitionistic Propositional
  Theorem Provers}.
\newblock In {\em Practical Aspects of Declarative Languages - 21th
  International Symposium, {PADL} 2019, Lisbon, Portugal, January 14-15, 2019,
  Proceedings}, {J.~J. Alferes} {and} {M.~Johansson}, Eds. Lecture Notes in
  Computer Science, vol. 11372. Springer, 115--132.

\bibitem[\protect\citeauthoryear{Tarau}{Tarau}{2020}]{lopstr20}
{\sc Tarau, P.} 2020.
\newblock {Synthesis of Modality Definitions and a Theorem Prover for Epistemic
  Intuitionistic Logic}.
\newblock In {\em Logic-Based Program Synthesis and Transformation - 30th
  International Symposium, {LOPSTR} 2020, Bologna, Italy, September 7-9, 2020,
  Proceedings}, {M.~Fern{\'{a}}ndez}, Ed. Lecture Notes in Computer Science,
  vol. 12561. Springer, 329--344.

\bibitem[\protect\citeauthoryear{Tarau and de~Paiva}{Tarau and
  de~Paiva}{2020}]{iclp20}
{\sc Tarau, P.} {\sc and} {\sc de~Paiva, V.} 2020.
\newblock {Deriving Theorems in Implicational Linear Logic, Declaratively}.
\newblock In {\em Proceedings, 36th International Conference on Logic
  Programming (Technical Communications)}, {F.~Ricca}, {A.~Russo}, {S.~Greco},
  {N.~Leone}, {A.~Artikis}, {G.~Friedrich}, {P.~Fodor}, {A.~Kimmig}, {F.~A.
  Lisi}, {M.~Maratea}, {A.~Mileo}, {and} {F.~Riguzzi}, Eds. {EPTCS}, vol. 325.
  110--123.

\end{thebibliography}

\end{document}